\documentclass[dvipsnames,sigconf=true, nonacm=true, review=false, anonymous = false,]{acmart}
\AtBeginDocument{%
  }

\setcopyright{acmlicensed}
\copyrightyear{2025}
\acmYear{2025}
\acmDOI{XXXXXXX.XXXXXXX}
\acmConference[Conference acronym 'XX]{Make sure to enter the correct
  conference title from your rights confirmation email}{June 03--05,
  2025}{Woodstock, NY}
\acmISBN{978-1-4503-XXXX-X/2018/06}

\usepackage{subcaption}
\usepackage{lipsum}
\usepackage{booktabs}
\usepackage{multirow}
\usepackage{framed,soul}
\usepackage{tabularx}
\usepackage{booktabs}
\usepackage{placeins}

\begin{document}

\title{BLADE: Benchmark suite for LLM-driven Automated Design and Evolution of iterative optimisation heuristics}


\author{Niki van Stein}
\email{n.van.stein@liacs.leidenuniv.nl}
\orcid{0000-0002-0013-7969}
\affiliation{%
  \institution{LIACS, Leiden University}
  \city{Leiden}
  \country{Netherlands}
  \postcode{2333 CC}
}

\author{Anna V. Kononova}
\email{a.kononova@liacs.leidenuniv.nl}
\orcid{0000-0002-4138-7024}
\affiliation{%
  \institution{LIACS, Leiden University}
  \city{Leiden}
  \country{Netherlands}
  \postcode{2333 CC}
}

\author{Haoran Yin}
\email{h.yin@liacs.leidenuniv.nl}
\orcid{0009-0005-7419-7488}
\affiliation{%
  \institution{LIACS, Leiden University}
  \city{Leiden}
  \country{Netherlands}
  \postcode{2333 CC}
}

\author{Thomas B{\"a}ck}
\email{t.h.w.baeck@liacs.leidenuniv.nl}
\orcid{0000-0001-6768-1478}
\affiliation{%
  \institution{LIACS, Leiden University}
  \city{Leiden}
  \country{Netherlands}
  \postcode{2333 CC}
}

\renewcommand{\shortauthors}{van Stein et al.}

\begin{abstract}
The application of Large Language Models (LLMs) for Automated Algorithm Discovery (AAD), particularly for optimisation heuristics, is an emerging field of research. This emergence necessitates robust, standardised benchmarking practices to rigorously evaluate the capabilities and limitations of LLM-driven AAD methods and the resulting generated algorithms, especially given the opacity of their design process and known issues with existing benchmarks. To address this need, we introduce BLADE (Benchmark suite for LLM-driven Automated Design and Evolution), a modular and extensible framework specifically designed for benchmarking LLM-driven AAD methods in a continuous black-box optimisation context. BLADE integrates collections of benchmark problems (including MA-BBOB and SBOX-COST among others) with instance generators and textual descriptions aimed at capability-focused testing, such as generalisation, specialisation and information exploitation. It offers flexible experimental setup options, standardised logging for reproducibility and fair comparison, incorporates methods for analysing the AAD process (e.g., Code Evolution Graphs and various visualisation approaches) and facilitates comparison against human-designed baselines through integration with established tools like IOHanalyser and IOHexplainer. BLADE provides an `out-of-the-box' solution to systematically evaluate LLM-driven AAD approaches. The framework is demonstrated through two distinct use cases exploring mutation prompt strategies and function specialisation.
\end{abstract}


\begin{teaserfigure}
  \includegraphics[width=\textwidth,trim=0mm 0mm 0mm 0mm,clip]{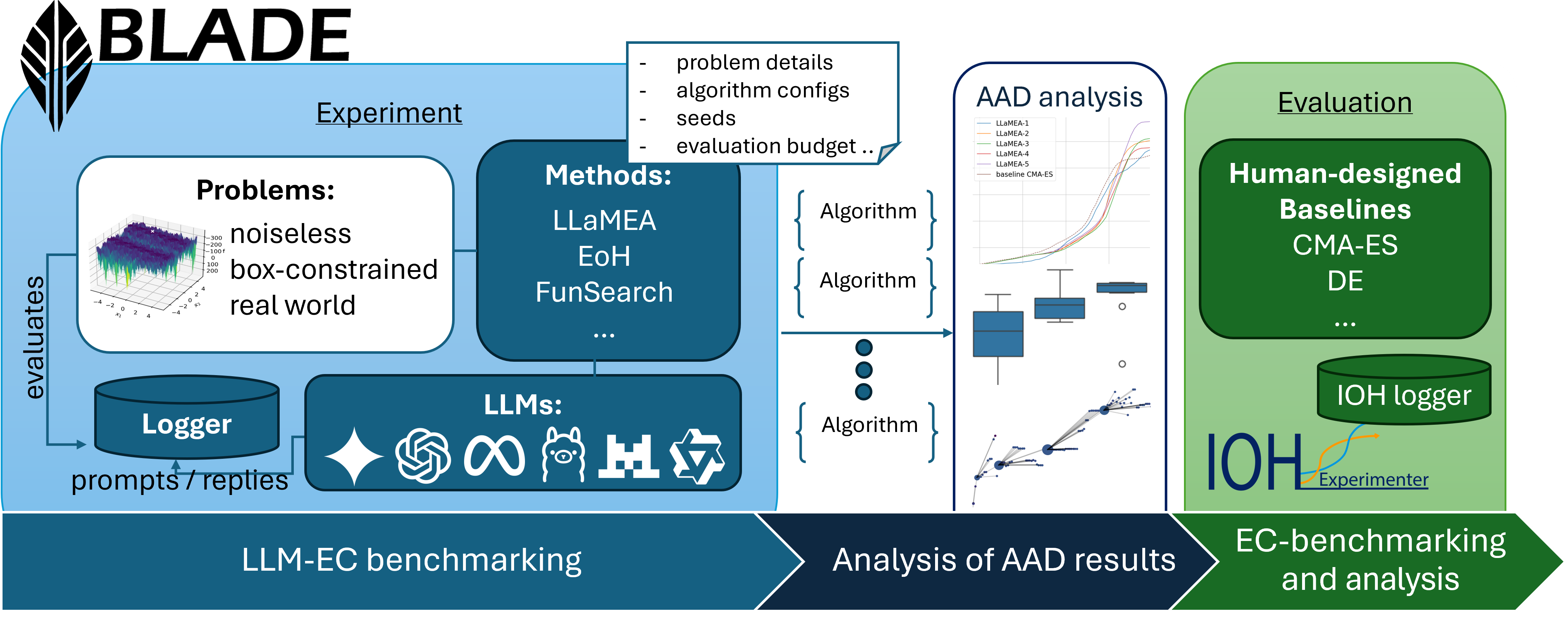}
  \caption{Abstract representation of the BLADE framework for benchmarking LLM-driven Automated Algorithm Discovery for continuous black-box optimisation.}
  \Description{Flow-diagram of the BLADE framework for benchmarking LLM-driven Automated Algorithm Discovery for continuous black-box optimisation.}
  \label{fig:teaser}
\end{teaserfigure}

\received{20 February 2007}
\received[revised]{12 March 2009}
\received[accepted]{5 June 2009}

\maketitle

\section{Introduction}
%

Large Language Models (LLMs) are increasingly being used as automated algorithm designers in optimisation~\cite{liu2024survey}. In this emerging paradigm, an LLM’s knowledge and reasoning ability are combined with traditional optimisation frameworks to discover new algorithms or improve existing ones. Recent studies have rapidly proliferated, showing that LLMs can serve as algorithm generators for various optimisation tasks~\cite{liu2024evolution,van2024llamea,SAT-LLM2024,liu2024llm4ad,sim2025beyond}. 

This field has become known as automatic algorithm discovery (AAD), whose goal is to have LLMs propose heuristic strategies or (snippets of) executable code for solving optimisation problems with minimal human input. It is widely accepted that the performance of optimisation algorithms depends on the optimisation problem, or, in other words, algorithms specialise on certain problem classes, following the result widely known as the No Free Lunch theorem~\cite{wolpert1995no}. Establishing the strengths and weaknesses of algorithms is a task of benchmarking, which has become a non-negotiable step in modern algorithm development, underpinned by established best practices~\cite{bartzbeielstein2020,Lopez2021Reproducibility}. More specifically, \textit{benchmarking} encompasses a systematic evaluation of algorithm performance on a curated set of problem instances, using defined metrics to assess and compare efficiency, accuracy and robustness across varying conditions or landscapes. Benchmarking becomes even more critical for automatically discovered algorithms, whose construction is opaque and often defies intuitive explanation.

It is essential to judiciously select the optimisation problems included in a benchmarking suite, as they must reflect the intended objectives of benchmarking: they should be numerous—without rendering the effort impractical—and offer \textit{sufficient diversity} within the problem class targeted by the algorithm. A similar rationale applies to the selection of performance measures, which should align with the benchmarking objectives. For this reason, a benchmarking framework with a \textit{modular design} is advisable, enabling components such as problem sets and performance measures to be easily swapped in or out depending on the goals of each experiment.

Unfortunately, concepts such as diversity and class of problems, which are essential for defining benchmarking setups in practice, remain elusive when it comes to rigorous formalisation. Nevertheless, researchers continue to invest considerable effort in this active area. Providing flexible tools that facilitate easy experimentation with benchmarking setups is therefore highly beneficial.

In this paper, we propose \textbf{BLADE: Benchmark suite for LLM-driven Automated Design and Evolution of iterative optimisation heuristics}, which serves the purpose. An overview of BLADE is presented in Figure~\ref{fig:teaser}. As demonstrated throughout the paper, BLADE includes:
\begin{enumerate}
    \item several collections of single-objective continuous optimisation problems with instance generation mechanisms, which are aimed at studying capabilities of algorithms such as, e.g., generalisation over problem instances and dimensionalities, levels of specialisation and efficiency of domain knowledge integration;
    \item flexible specification of benchmarking setup used inside AAD, such as, e.g., run budgets, number of runs and instances used for training and testing (allowing to evolve the algorithm on one set of problem instances and then test on a novel instance to truly assess generalisation); 
    \item a standardized interface to connect with many different LLMs and postprocess the generated output to get code and descriptions of solutions;
    \item methods for the analysis of the results of the algorithm discovery process, such as, e.g., Code Evolution graphs~\cite{vanstein2025codeevolutiongraphsunderstanding} and ELO ratings~\cite{albers2001elo};
    \item methods for performance comparison of discovered algorithms against human-designed or other specified algorithms via native integration with tools like IOHanalyser~\cite{IOHanalyzer} and IOHxplainer~\cite{IOHexplainer}. 
\end{enumerate}

Thanks to BLADE’s modular design, all these components can be readily extended. At the same time, BLADE works `out-of-the-box', using loggers that record all necessary information by default, which is then passed to the specified AAD and evolutionary computation benchmarking analysis tools. The current focus on continuous black-box optimisation stems both from our prior expertise and from a clear gap in the existing literature.

This paper is organised as follows: Section~\ref{sect:related} summarises relevant related work, Section~\ref{sect:methodology} outlines full paper methodology, Section~\ref{sect:use-cases} discusses two use-cases selected to demonstrate the usage of BLADE, Section~\ref{sect:repro} provides a reproducibility and transparency statement and Section~\ref{sect:conclusions} concludes with final remarks and future directions.

\section{Related work}\label{sect:related} 

\subsubsection*{From AS to AAD to AAD benchmarking} 
Automatic algorithm discovery is rooted in algorithm selection (AS) approaches, which initially aimed to identify well-performing algorithms from a given portfolio of (parameterised) algorithms for a specific problem instance. Traditional approaches to AS relied on performance prediction models, often informed by exploratory landscape features and used tools such as irace~\cite{Lopez2016irace} and SMAC~\cite{SMAC3}. The introduction of modular frameworks such as modular CMA-ES (modCMA)~\cite{modcma} and modular Differential Evolution (modDE)~\cite{modDE} extended these methods by enabling fine-grained configuration of algorithmic components. 

While these tools significantly advanced the state of the art, they still required substantial expert input, both in defining the portfolio of candidate algorithms and in engineering suitable instance features. With the advent of powerful Large Language Models, the field has begun shifting from selecting among pre-existing algorithms to automatically generating novel algorithms—an approach now commonly referred to as Automatic Algorithm Discovery.


In this emerging paradigm, LLMs are prompted to propose algorithmic strategies, often expressed as pseudocode or directly executable code, that are then evaluated using traditional optimisation benchmarks. AAD frameworks such as FunSearch~\cite{FunSearch2024}, LLaMEA~\cite{van2024llamea} and the Evolution of Heuristics (EoH) framework~\cite{liu2024evolution} exemplify this direction. These systems enable LLMs to iteratively improve their outputs by incorporating feedback from fitness evaluations. The LLaMEA framework, in particular, has seen several recent extensions, including variants that incorporate hyper-parameter optimisation~\cite{vanstein2024hpo}, mutation control~\cite{yin2024controlling} and real-world problem specialisation~\cite{yin2025optimizingphotonicstructureslarge}.

A significant body of novel research on AAD~\cite{liu2024survey} brought the need for a platform facilitating algorithm discovery using large language models. LLM4AD~\cite{liu2024llm4ad} has thus been proposed as a broad modular meta-framework designed to support the general development of LLM-driven algorithm discovery methods in a standardised evaluation environment. LLM4AD is primarily focused on combinatorial optimisation and is in its early development stage.



\subsubsection*{Issues with existing benchmarking setups} 
The de-facto standard for continuous optimisation is the BBOB (Black-Box optimisation Benchmark) suite~\cite{bbob_hansen2009_noiseless}, part of the COCO platform~\cite{coco_hansen2021}, which provides a set of 24 functions defined for any dimensionality, with varying properties (unimodal to highly multimodal, separable vs. non-separable, etc.) and allows generating multiple instances per function (by shifting or rotating the coordinate system). By using BBOB, researchers ensure generality in evaluation – an algorithm must handle rugged, smooth, separable and non-separable problems, rather than being overfitted to one test function.

While representing a definitive step forward in advancing benchmark methodology, BBOB comes with a range of problems:
\begin{itemize}
    \item Ambiguous formulation of the optimisation problem (unconstrained vs box-constrained) and insufficient diversity of problem instances in terms of locations of optima within the search domain of BBOB problems \cite{bbobinstancealanysis}. To remedy this, SBOX-COST function suite has been proposed \cite{vermetten2023analysis}. 
    \item Models trained on BBOB data generalise poorly to other suites such as those provided by the CEC conference~\cite{cec2013,cec2014,cec2015} due to statistically significant differences in feature-space distributions~\cite{nikolikj2024}. This stems from the unintended interpretation of the generality of functions included in BBOB.
    \item Similarly, MA-BBOB~\cite{Vermetten2025MABBOB} which extends BBOB by generating a wide array of benchmark functions through affine combinations of BBOB functions, the resulting problem landscapes may not exhibit sufficient diversity~\cite{Dietrich2024}. This limitation poses challenges for automated algorithm selection methods, as training on these generated functions does not necessarily lead to effective generalisation to novel problem instances. The authors emphasize the need for careful selection of training instances to ensure robust and generalizable algorithm selection models~\cite{Dietrich2024}.
\end{itemize}

Furthermore, results in combinatorial optimisation~\cite{sim2025beyond} demonstrate that most heuristics discovered via LLMs also fail to generalise across diverse problem instances, performing well only in specific areas of the instance space. In contrast, traditional simple heuristics demonstrate in this case more consistent performance across a broader range of benchmarks.  Thus, the authors advocate for the development and utilisation of more diverse benchmarks and applications to gain a better understanding of this emerging paradigm in AAD~\cite{sim2025beyond}. Performance analysis on the results of AAD also suggests that benchmark suites for bin packing problems are simply too easy \cite{sim2025beyond,vanstein2024hpo}. 


\section{Methodology}\label{sect:methodology}
This section details the methodology of BLADE for benchmarking LLM-driven Automated Algorithm Discovery. 
The framework, as depicted in Figure 1, encompasses several key components: benchmark problems, AAD methods, LLMs and evaluation metrics.
The BLADE framework uses the following key steps in the benchmarking pipeline:

\subsubsection*{Experimental setup} 
First, an experiment is designed, or one of the provided experimental setups is used directly for easy comparison with other works. An experiment consists of a list of search methods (such as LLaMEA, EoH, FunSearch, Random Search, ReEvo and others), a list of optimisation problems and an LLM. In addition, the experiment has several attributes such as the number of algorithms per run to evaluate (AAD budget), the number of independent runs per method/problem combination and some other settings. All experiment settings are logged for reproducibility.

\subsubsection*{LLM-EC Benchmarking} 
Next, the experiment(s) is/are carried out using the experimental setup provided. BLADE uses parallel processes to speed up the evaluations of the different runs over different CPU threads. \textit{Loggers} are attached to the LLM and Problem objects to log all LLM queries made and all generated algorithm evaluations. The loggers operate independently of the search method and, as such, enforce a fair comparison between LLM-EC (search) algorithms. The LLM token costs, seeds and other hyperparameters are logged for transparency and reproducibility. On the optimisation problem side, each problem evaluation is logged. Problems consist of several (predefined) training and testing instances, where the training instances are used during the LLM-EC runs and the test instances are used for the final evaluation of best best-performing generated algorithms.

\subsubsection*{Analysis of AAD results} 
After the different LLM-EC runs, the convergence of the different algorithms can be analysed by visualising the best-so-far convergence curves averaged over all random seeds. In addition, we can analyse in more detail how the code evolved in one or more of the runs by looking at the Code Evolution Graphs (CEG)~\cite{vanstein2025codeevolutiongraphsunderstanding}. The CEG shows a low-dimensional embedding of various static code features against the number of algorithms generated. 

\subsubsection*{EC Benchmarking and Analysis} 
The final step in the BLADE benchmarking toolbox is the evaluation of the best found algorithms on separate \textit{validation instances} of the problem. These instances can either be larger versions of real-world problems or different instances of the same optimisation problems that the LLM-EC has used during the LLM-EC benchmarking phase. The best algorithms found by each LLM-driven AAD method are compared against each other and also against the state-of-the-art human-designed algorithms such as CMA-ES~\cite{hansen2003reducing}, Differential Evolution variants~\cite{feoktistov2006differential} and others.
The final analysis of the comparison depends on the problem specification. For example, we can look at the optimal fitness reached per algorithm or at anytime performance metrics such as the Area Over the Convergence Curve (AOCC)~\cite{lopez2024using} and the Empirical Attainment Function (EAF)~\cite{lopez2024using}, which give better insight into the effectiveness of the different algorithms over different evaluation budgets. For the visualisation and logging of the final evaluation, BLADE is integrated with IOHanalyser~\cite{IOHanalyzer} and IOHexplainer~\cite{IOHexplainer}, the established benchmarking tools of the black-box optimisation field.

\subsection{Capability-Focused Benchmarking}\label{sec:capability_benchmarking}
To move beyond generic black-box optimisation evaluations, focusing purely on benchmarking performance and addressing the limitations highlighted by the No Free Lunch theorem~\cite{wolpert1995no}, we advocate for \textit{capability-focused benchmarking}. This approach allows for targeted experiments designed to answer specific research questions regarding the performance of AAD methods, particularly those driven by LLMs. We aim to assess the following key capabilities:

\begin{itemize}
    \item \textbf{Generalization:} Evaluating an AAD method's ability to generate optimisation algorithms effective across a diverse range of problems within a given dimensionality.
    \item \textbf{Problem Class Specialization:} Assessing an AAD method's proficiency in evolving solvers tailored for specific classes of problems, e.g., such as those characterised by multi-modality with weak global structure.
    \item \textbf{Information Exploitation:} Determining an AAD method's capacity to leverage problem-specific details (e.g., textual descriptions of landscape features) to construct specialised solvers for particular problem instances.
\end{itemize}

These capabilities reflect the potential of LLM-EC methods to produce solvers ranging from general-purpose tools to highly specialised algorithms for specific problem domains or individual problems (per-instance configuration). To rigorously test these capabilities, we propose utilising the following benchmark suites:

\begin{itemize}
    \item \textbf{Many Affine Black-Box optimisation Benchmark (MA-BBOB):} This configurable suite comprises functions derived from BBOB~\cite{bbob_hansen2009_noiseless}, originally proposed as part of the comparing continuous optimizers (COCO) environment~\cite{coco_hansen2021}, via affine transformations~\cite{vermetten2024ma}. We recommend using a predefined set of $20$ training and $50$ testing instances and provide in total $1000$ predefined instances. A key feature of MA-BBOB is the random distribution of optima within the search domain, mitigating the risk of LLMs succeeding through memorisation or biased initialisation strategies~\cite{cmaes_vermetten2022} (a known issue in some of the original BBOB functions \cite{bbobinstancealanysis}). MA-BBOB is primarily used to assess generalisation capabilities.
    \item \textbf{SBOX-COST:} This benchmark suite features box-constrained black-box optimisation problems \cite{vermetten2023analysis}, also with randomly distributed optima across the full domain. It includes five distinct function groups, each equipped with an instance generator (applying rotations and shifts). We included textual descriptions of the underlying optimisation problem landscapes. SBOX-COST is employed for evaluating group-level and function-level specialization, including the ability to utilize provided problem (or problem class) information.
\end{itemize}

These synthetic benchmarking suites have been chosen because of their recognition in the field (they are based on the popular BBOB suite, but without the center bias of the location of optima) and because they are designed to test different capabilities in optimisation algorithms. These capabilities are for example, how well the algorithms can escape local optima, how well they can exploit global structure, etc.

\subsection{Real world applications}

Next to the synthetic benchmarking functions that are used to test various capabilities in LLM-driven AAD methods, several \textit{real-world problems} are included (and more will be added) to BLADE to verify the capabilities of AAD methods in practical domains.

In ~\cite{yin2025optimizingphotonicstructureslarge}, LLaMEA is used to solve real-world photonic structure optimisation problems from \cite{bennet2024illustrated}, including Bragg mirror problems, inverse ellipsometry problems and photovoltaic design problems, which have practical implications for communication, semiconductors, LED displays, materials analysis and solar cells~\cite{yin2025optimizingphotonicstructureslarge}. 

Since the evaluation of these (and many other) real-world problems is very time-consuming, the AAD methods should use smaller-scale versions of these problems (with a lower dimensionality) as \textit{training instances} and \textit{validate} the best found algorithms on the high-dimensional harder problems.
In addition, background knowledge on the real-world problems is typically available (from physics) to be used inside the prompt by the LLM-driven AAD methods. 
Initial results in \cite{yin2025optimizingphotonicstructureslarge} confirmed that this knowledge usually has a positive effect on the quality of the algorithms generated, however, this likely depends on the LLM and on the real-world problem and requires deeper investigation.

\section{Use-cases}\label{sect:use-cases}
To effectively show how the proposed BLADE toolbox works, we include two use-cases. The first use-case compares different mutation prompting strategies and uses analysis on static code features to discover the effect of different prompting techniques. In addition, it shows the BLADE benchmarking pipeline step-by-step. The second use-case shows how to test for function specialisation where the LLMs can exploit additional information about the problems to solve given in the task prompt for different types of problems. The first use-case focuses more on the EC part of the LLM-driven AAD process, while the latter use-case focuses more on LLM capabilities, showing two (of many) different approaches possible with the BLADE framework.

\subsection{Use-case: Mutation Prompts}\label{sec:use-case_mutation}

\begin{figure}[!t]
    \includegraphics[width=0.8\linewidth,trim=2mm 3mm 2mm 8mm,clip]{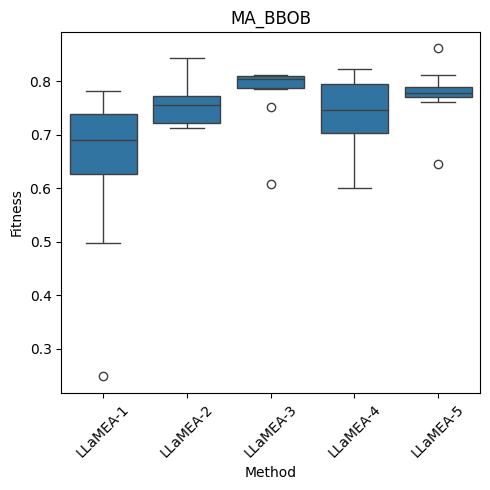}
    \caption{Distribution of the fitness (AOCC) of algorithms resulting from each AAD run. The AOCC is obtained by evaluating each of the resulting algorithms ($10$ algorithms per LLaMEA configuration) on $50$ evaluation instances of MA-BBOB and $10$ random seeds. \label{fig:mutation_aad_fitness}}
\end{figure}
This use-case investigates the impact of different mutation prompt strategies on the effectiveness of LLM-driven algorithm discovery using the LLaMEA \cite{van2024llamea} framework with a (4, 12) evolution strategy (4 parents, 12 offsprings). 

The central research question addressed is: \emph{Which combination of mutation prompts is most efficient for generating general-purpose continuous black-box optimizers when using a specific LLM (Gemini-2.0-flash)?}

\subsubsection{Experimental Setup}
The experiment was conducted using the $5$-dimensional MA-BBOB benchmark suite~\cite{vermetten2024ma}, chosen for its diverse set of problems with randomly distributed optima. The LLM-EC process utilised $20$ predefined MA-BBOB training instances, while the final evaluation was performed on $50$ distinct test instances. Each LLM-EC run had a budget of generating (and evaluating) $100$ candidate algorithms and each algorithm evaluation within the EC process had a budget of $2000 \times \text{dimensionality} = 10,000$ function evaluations.

Three distinct mutation prompts were defined:
\begin{enumerate}
    \item "Refine the strategy of the selected algorithm to improve it"
    \item "Generate a new algorithm that is different from the algorithms you have tried before"
    \item "Refine and simplify the selected algorithm to improve it"
\end{enumerate}

Five LLaMEA configurations were tested, each employing different combinations of these mutation prompts:
\begin{itemize}
    \item LLaMEA-1: Uses only prompt 1
    \item LLaMEA-2: Uses only prompt 2
    \item LLaMEA-3: Uses only prompt 3
    \item LLaMEA-4: Uses prompts 1 and 2 (chosen randomly)
    \item LLaMEA-5: Uses all three prompts (chosen randomly)
\end{itemize}

\subsubsection{AAD Evaluation Results}
\begin{figure}[!tb]
    \includegraphics[width=\linewidth,trim=4mm 4mm 4mm 4mm,clip]{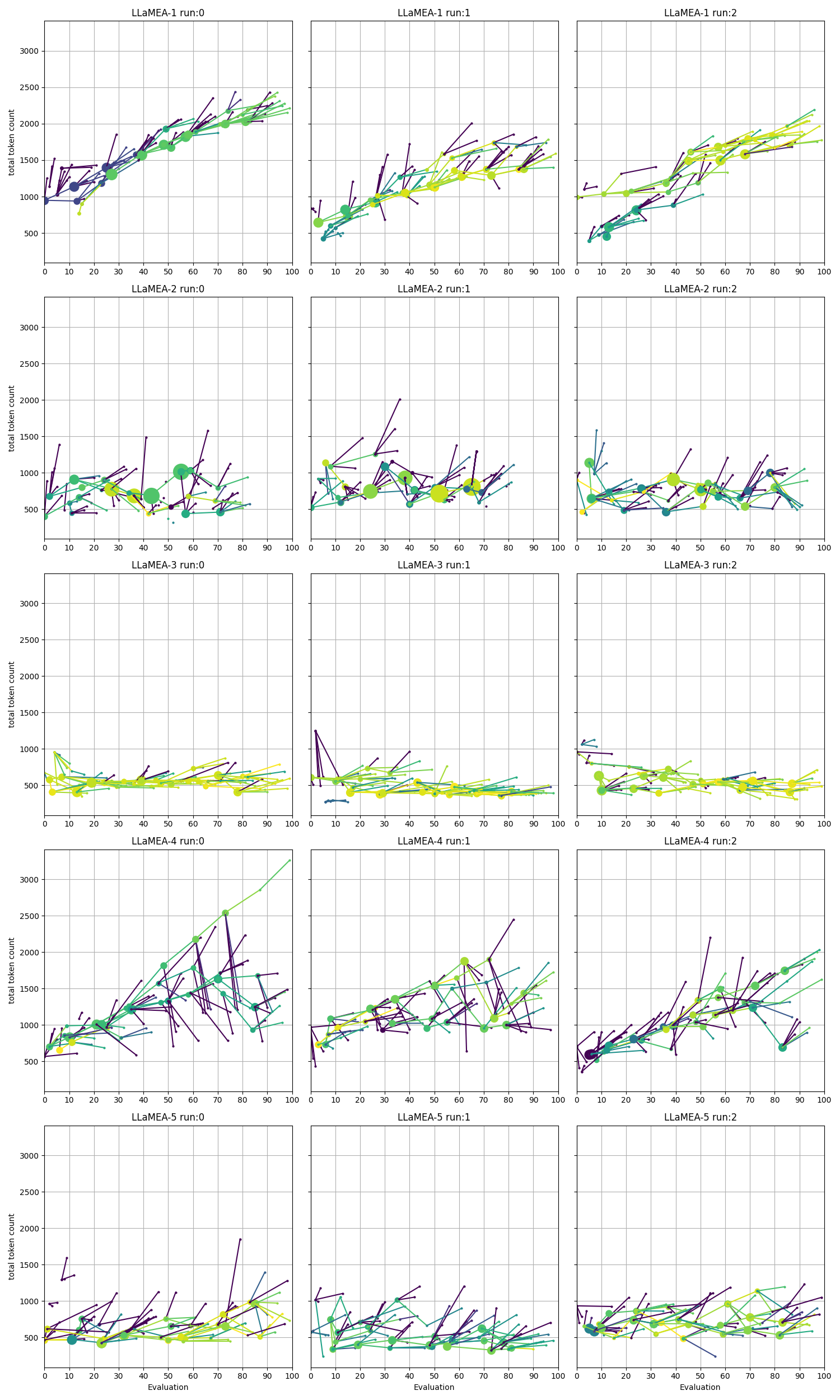}
    \caption{Code Evolution Graphs of $3$ runs (columns) per LLaMEA variant (rows) showing how the number of Python tokens (y-axis) changes per generated algorithm (x-axis). The colour of the nodes denotes the AOCC score (yellow is higher and better, blue is lower and worse), while connections between nodes denote parent-offspring relations and the size of nodes encodes the number of connections produced from this node. Runs are sometimes showing fewer than $100$ evaluations due to runtime errors inside the generated algorithms (resulting in a negative fitness which is not displayed). \label{fig:CEG}}
\end{figure}

Figure~\ref{fig:mutation_aad_fitness} illustrates the distribution of fitness values, measured by the Area Over the Convergence Curve (AOCC), for the algorithms produced by each LLaMEA configuration during the automated design phase. Each boxplot represents the AOCC values obtained by evaluating the $10$ algorithms generated (by $10$ different runs) of each respective LLaMEA variant on the $50$ MA-BBOB evaluation instances, aggregated over $10$ independent seeds ($500$ evaluation runs per algorithm). Visual inspection suggests variability in the effectiveness of different prompt strategies, with configurations using prompt 3 (``simplify'') (LLaMEA-3 and LLaMEA-5) showing potential for generating higher-performing algorithms compared to the default prompt 1 ``refine'' option provided by vanilla LLaMEA. In addition, when we look at the Code Evolution Graphs in Figure \ref{fig:CEG}, we can clearly observe that including prompt 3 (LLaMEA-3 and 5) keeps the size of the code small. It can also be observed that adding prompt 2 (LLaMEA-2, 4 and 5) increases the diversity in code substantially.

\subsubsection{Validation Against Human-Designed Baseline}
To assess the quality of the generated algorithms relative to established methods, the best-performing algorithm discovered by each of the five LLaMEA configurations (selected based on performance across 10 independent AAD runs and their training instance performance) was further evaluated. These selected algorithms were compared against CMA-ES, a state-of-the-art human-designed evolutionary algorithm, serving as a strong baseline~\cite{hansen2003reducing}. The evaluation involved running each selected algorithm and the baseline on the $50$ MA-BBOB test instances over $10$ independent runs.

\begin{figure}[!t]
    \includegraphics[width=\linewidth,trim=3mm 4mm 1mm 3mm,clip]{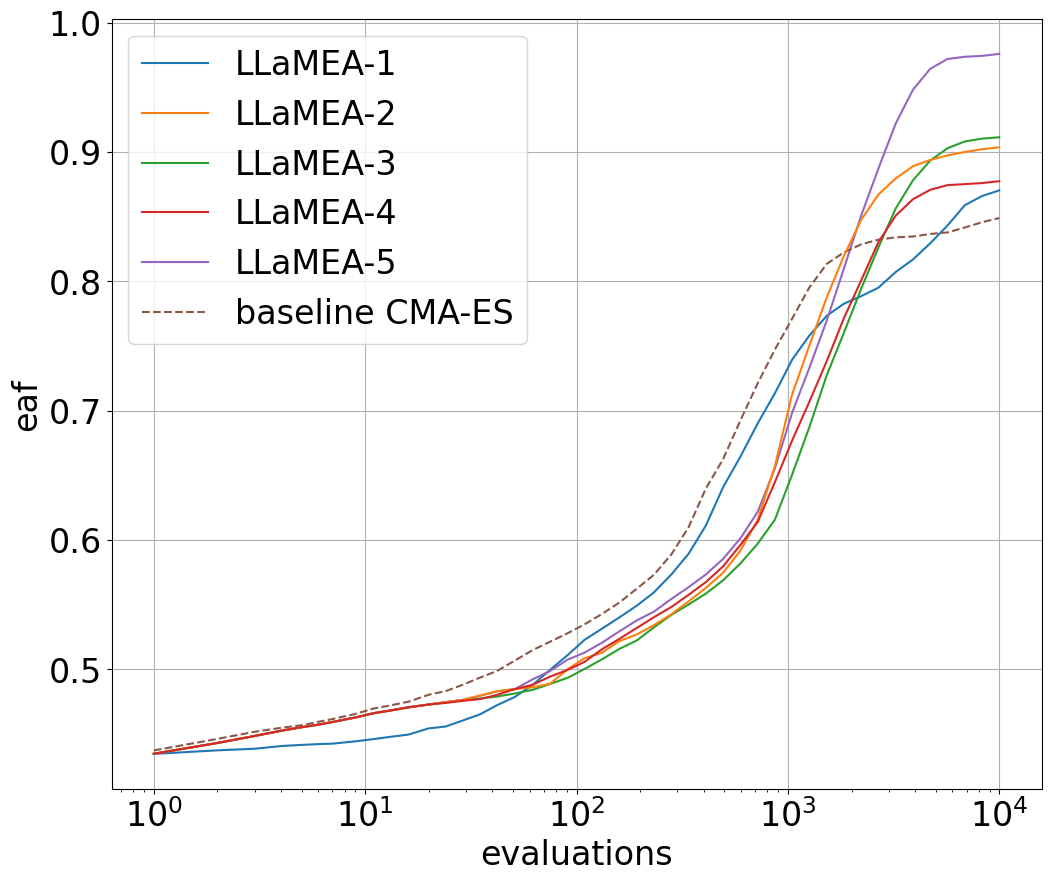}
    \caption{Empirical Attainment Function (EAF) curves of the best solutions found per LLaMEA configuration (best of $10$ runs) and a CMA-ES baseline. Each algorithm is tested on $50$ different MA-BBOB instances and $10$ independent runs. \label{fig:mutation_eaf}}
\end{figure}

The performance comparison is visualised using Empirical Attainment Function (EAF) curves and ELO ratings~\cite{albers2001elo}. Figure~\ref{fig:mutation_eaf} displays the EAF curves, illustrating the probability of reaching a certain target function value within a given budget of function evaluations~\cite{lopez2024using}. This provides insight into the anytime performance characteristics of the algorithms. The EAF curves show that the best algorithms generated by several LLaMEA configurations, particularly LLaMEA-5, are competitive with or outperform the CMA-ES baseline, especially in the later stages of the search corresponding to higher evaluation budgets.

\begin{figure}[!t]
    \includegraphics[width=\linewidth,trim=2mm 4mm 2mm 2mm,clip]{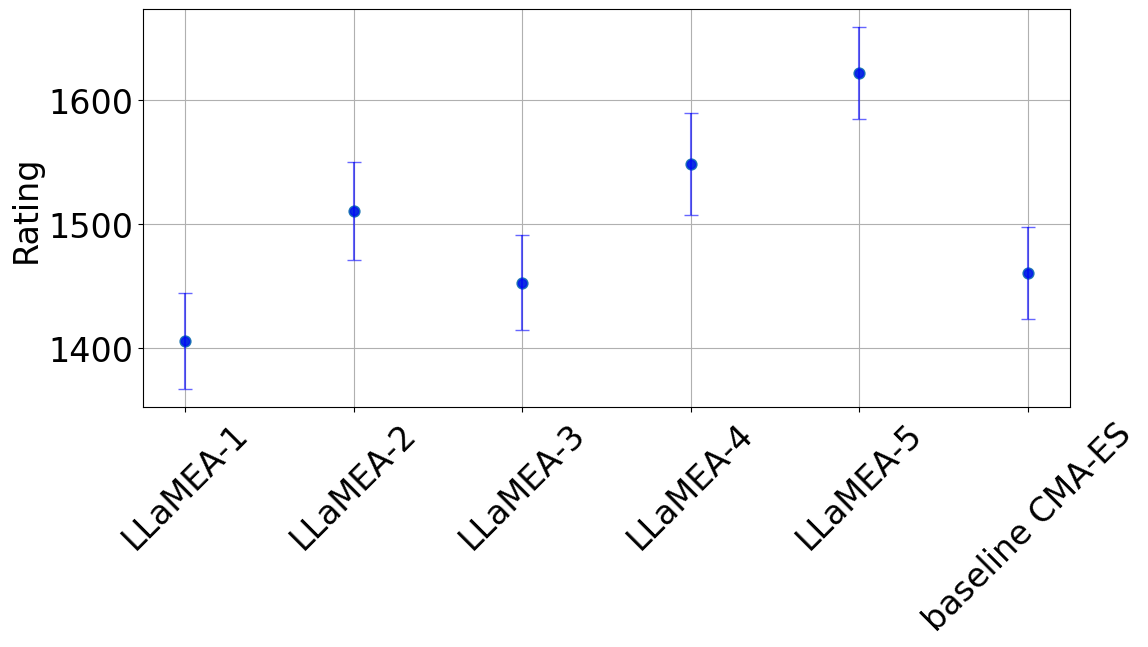}
    \caption{ELO rating (higher is better) of the best solutions found per LLaMEA configuration (best of $10$ runs) and a CMA-ES baseline. Each algorithm is tested on $50$ different MA-BBOB instances and $10$ independent runs. ELO rating is calculated using a tournament of $100\,000$ one against one comparisons. \label{fig:mutation_elo}}
\end{figure}

Figure~\ref{fig:mutation_elo} presents the ELO ratings calculated from pairwise comparisons between the best algorithm from each LLaMEA configuration and the CMA-ES baseline. The ELO rating provides a single metric summarising the relative strength of each algorithm based on tournament-style comparisons across the $50$ test instances and $10$ runs using a tournament size of $100\,000$ pair-wise comparisons. The results indicate that LLaMEA-5 achieved the highest ELO rating, suggesting that utilising a combination of all three mutation prompts yielded the most robust algorithm overall within this experimental setup, outperforming other configurations and the CMA-ES baseline. LLaMEA-2 and LLaMEA-4 also demonstrate strong performance relative to the baseline according to this metric.

\subsection{Use-case: LLM Comparison}
This use-case aims to evaluate and compare the effectiveness of different LLMs in the context of AAD for black-box optimisation. Specifically, it seeks to determine \textit{which LLM is most proficient in generating optimizers tailored to a specific type of function or problem class, focusing on specialisation in landscape function properties}.

The experiments involve using LLaMEA with the following LLMs: local models codestral (22b) and  qwen2.5-coder (14b) and closed-source models gemini-1.5-flash \cite{geminiteam2024gemini15unlockingmultimodal} and the more recently introduced gemini-2.0-flash \footnote{Specifically codestral \url{https://ollama.com/library/codestral:22b} and qwen2.5-coder \url{https://ollama.com/library/qwen2.5-coder:14b} from the ollama repository.}. 

The experiment is comprised of two parts. The per-function specialisation and the per-function-group specialisation. In the \textit{per-function} specialisation part, the AAD methods are given $5$ instances of a particular optimisation problem including a textual description of the problem, such as ``Separable Ellipsoidal Function'' in the case of function 2 ($fid2$). The best resulting algorithm per run is then evaluated using $10$ (different) evaluation instances of the same problem.
This is repeated for different functions (with different characteristics), in this case we choose fid2 (Separable Ellipsoidal Function), fid5 (Linear Slope), fid13 (Sharp Ridge Function), fid15 (Rastrigin Function) and fid21 (Gallagher's Gaussian 101-me Peaks Function) as these functions form a very diverse set. In the \textit{per-function-group} part of the experiment, the LLM-driven AAD method is given $5$ instances of all problems in a particular function group (4 or 5 problems per group). The groups are: group1 "Separable Functions", group2 "Functions with low or moderate conditioning", group3 "Functions with high conditioning and unimodal",
group4 "Multi-modal functions with adequate global structure" and
group5 "Multi-modal functions with weak global structure".
The final algorithms are validated on $10$ different instances for each problem in the same group.

\begin{figure*}[!t]
    \includegraphics[width=\linewidth,trim=2mm 3mm 2mm 2mm,clip]{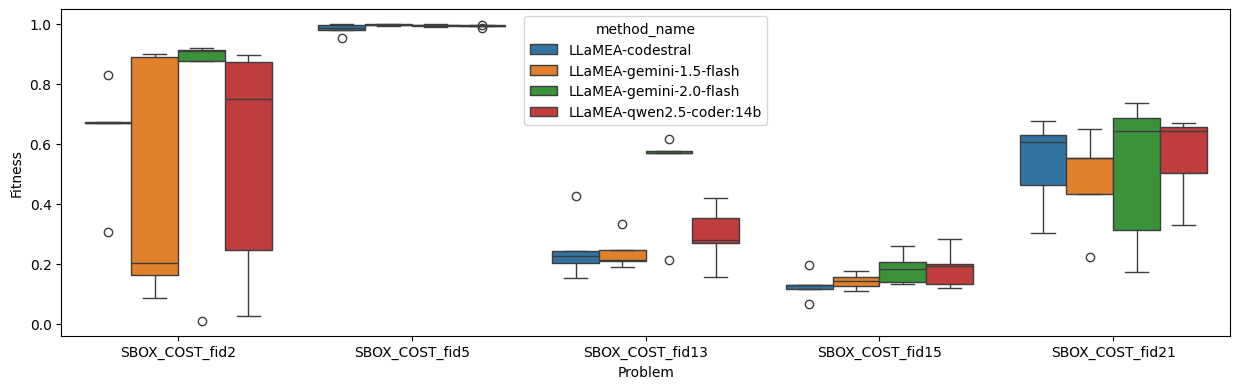}
    \includegraphics[width=\linewidth,trim=2mm 3mm 2mm 2mm,clip]{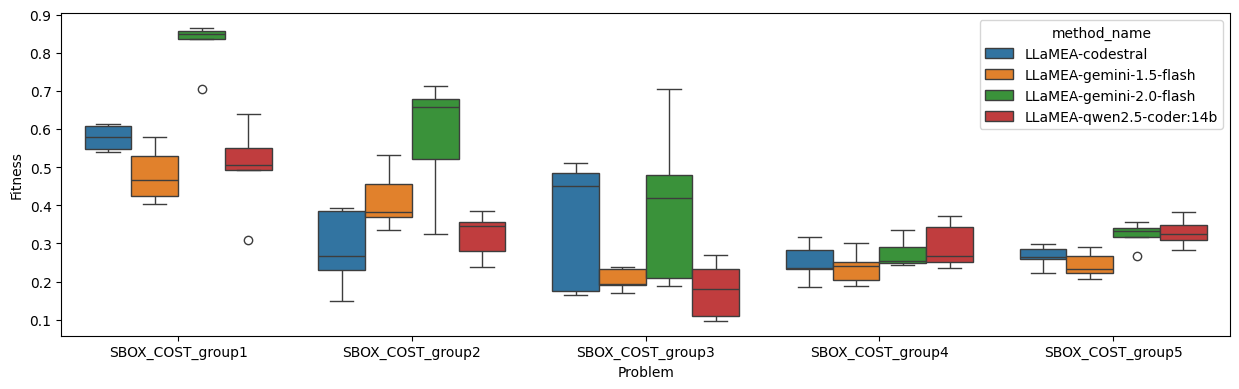}
    \caption{Fitness (AOCC) distribution of all final solutions generated using LLaMEA with different LLMs. \label{fig:SBOX}}
\end{figure*}

For both parts of the experiment, each LLM-EC run had a budget of generating $50$ candidate algorithms, and each algorithm evaluation within the EC process had a budget of $2000 \times d (5) = 10,000$ function evaluations. The performance of these LLMs in combination with the LLaMEA framework is assessed based on the fitness (AOCC) of the final solutions they generate, as illustrated in Figure \ref{fig:SBOX} and Table \ref{tab:AOCC} (higher is better). These results are derived from the validation instances, providing insight into the generalisation capabilities of the algorithms produced by each LLM.

\begin{table}[!t]
\centering
\caption{Mean AOCC for each LLM-EC over the final solutions of all runs per problem\label{tab:AOCC}. Results in boldface are significantly better with a threshold of $0.05$, p-values below this threshold are reported.}
\resizebox{\linewidth}{!}{%
\begin{tabular}{llllll}
\toprule
 & $f_2$ & $f_5$ &$ f_{13}$ & $f_{15}$ & $f_{21}$ \\
\midrule
codestral & 0.63 ± 0.17 & 0.98 ± 0.02 & 0.25 ± 0.09 & 0.13 ± 0.04 & 0.53 ± 0.14 \\
gemini-1.5-flash & 0.45 ± 0.37 & 1.00 ± 0.00 & 0.24 ± 0.05 & 0.14 ± 0.02 & 0.48 ± 0.15 \\
gemini-2.0-flash & 0.72 ± 0.36 & 0.99 ± 0.00 & \textbf{0.51 ± 0.15} (p=0.047) & 0.18 ± 0.05 & 0.51 ± 0.22 \\
qwen2.5-coder:14b & 0.56 ± 0.36 & 0.99 ± 0.00 & 0.29 ± 0.09 & 0.18 ± 0.06 & 0.56 ± 0.13 \\
\bottomrule
\end{tabular}}

\resizebox{\linewidth}{!}{%
\begin{tabular}{llllll}
\toprule
 & Group 1 & Group 2 & Group 3 & Group 4 & Group 5 \\
\midrule
codestral & 0.58 ± 0.03 & 0.29 ± 0.09 & 0.36 ± 0.15 & 0.25 ± 0.05 & 0.27 ± 0.03 \\
gemini-1.5-flash & 0.48 ± 0.07 & 0.41 ± 0.07 & 0.21 ± 0.03 & 0.24 ± 0.04 & 0.24 ± 0.03 \\
gemini-2.0-flash & \textbf{0.82 ± 0.06} (p=0.002) & 0.58 ± 0.14 & 0.40 ± 0.19 & 0.27 ± 0.03 & 0.32 ± 0.03 \\
qwen2.5-coder:14b & 0.50 ± 0.11 & 0.32 ± 0.05 & 0.18 ± 0.07 & 0.29 ± 0.05 & 0.33 ± 0.03 \\
\bottomrule
\end{tabular}}

\end{table}

In Figure \ref{fig:SBOX} we can observe that gemini-2.0-flash is outperforming other LLMs mostly for unimodal and low conditioning functions (fid2, group1, fid13, group2).
Other models show relatively similar performance and it is interesting to note that the \textit{smaller open-source models do not underperform significantly} in most cases. From Table \ref{tab:AOCC} we can further see that gemini-2.0-flash is only significantly better (using a pairwise independent t-test) in two of the ten problem setups.

\section{Reproducibility and Transparency}\label{sect:repro}
\textbf{BLADE} is available open-source and fully documented on Github~\footnote{\url{https://github.com/XAI-liacs/BLADE}}. All code and results of the experiments in this paper, including full prompts, generated algorithms and seeds are available in our Zenodo repository \cite{van_stein_2025_15119985}.

\section{Conclusions and Outlook}\label{sect:conclusions}
This paper introduces \textbf{BLADE}, a novel benchmarking suite designed to evaluate and compare LLM-driven Automated Algorithm Discovery (AAD) methods for continuous black-box optimisation. BLADE provides a structured framework encompassing various black-box benchmark problem sets, including real-world applications and synthetic suites tailored for capability-focused benchmarking. The \textit{modular design} of BLADE facilitates the integration of different AAD methods, LLMs and evaluation metrics, enabling comprehensive and reproducible experiments.

The use-cases presented in this paper demonstrate the utility of BLADE in analysing the impact of different prompt strategies on LLM-driven algorithm generation and in comparing the performance of various LLMs within the AAD context. The results highlight the potential of LLM-driven AAD to produce competitive optimizers while also revealing the influence of prompt engineering and LLM selection on the effectiveness of the generated algorithms.

Future work will focus on expanding the benchmark suite with additional real-world problems to assess other relevant capabilities of LLM-driven AAD methods. Furthermore, we plan to incorporate a wider range of baseline algorithms and explore automated hyperparameter optimisation within the benchmarking framework.

\FloatBarrier

\bibliographystyle{ACM-Reference-Format}
\bibliography{sample-base}

\appendix

\end{document}